\documentclass[twocolumn,showpacs,superscriptaddress,preprintnumbers,amsmath,amssymb,prc]{revtex4}

\usepackage{graphicx}
\usepackage{dcolumn}
\usepackage{bm}
\usepackage{float}
\usepackage{color}
\usepackage{CJK}
\usepackage{ulem}
\usepackage[colorlinks,linkcolor=blue,urlcolor=blue,citecolor=blue]{hyperref}

\begin{document}
\begin{CJK*}{UTF8}{gbsn}

\title{Green-Kubo formula for Boltzmann and Fermi-Dirac statistics}
\author{X. G. Deng (邓先概)}
\affiliation{Key Laboratory of Nuclear Physics and Ion-beam Application (MOE), Institute of Modern Physics, Fudan University, Shanghai 200433, China}
\affiliation{Shanghai Institute of Applied Physics, Chinese Academy
of Sciences, Shanghai 201800, China}
\affiliation{University of Chinese Academy of Sciences, Beijing 100049, China}

\author{Y. G. Ma (马余刚) \footnote{Corresponding author: mayugang@fudan.edu.cn}}
\affiliation{Key Laboratory of Nuclear Physics and Ion-beam Application (MOE), Institute of Modern Physics, Fudan University, Shanghai 200433, China}
\affiliation{Shanghai Institute of Applied Physics, Chinese Academy of Sciences, Shanghai 201800, China}

\author{Y. X. Zhang (张英逊) }
\affiliation{China Institute of Atomic Energy, Beijing 102413, China}

\date{\today}

\begin{abstract}

Shear viscosity of nuclear matter is extracted via the Green-Kubo formula and the Gaussian thermostated SLLOD algorithm (the shear rate method) in a periodic box by using an improved quantum molecular dynamic (ImQMD) model without mean field, also it is  calculated by a Boltzmann-type equation.  Here a new form of the Green-Kubo formula is put forward in the present work.  For classical limit at nuclear matter densities of $0.4\rho_{0}$ and $1.0\rho_{0}$, shear viscosity by the traditional and new form of the Green-Kubo formula as well as the SLLOD algorithm are coincident with each other. However, for non-classical limit, shear viscosity by the traditional form of the Green-Kubo formula is higher than those obtained by the new form of the Green-Kubo formula as well as the SLLOD algorithm especially in low temperature region. In addition, shear viscosity from  the Boltzmann-type equation is found to be less than that by the Green-Kubo method or the SLLOD algorithm for both classical and non-classical limits. 

\end{abstract}


\maketitle

\section{Introduction}
\label{introduction}

Shear viscosity is a common transport property for lots of substances, such as macroscopic matter, eg. water, oil, honey and air as well as microscopic and quantum matter, eg. hot dense quark matter etc.
\cite{Rev1,Rev2,Rev3,Gao,Tang,Huang,Shen,Cao,KEE15}, 
and it has been studied  for a few decades in nuclear physics \cite{GB78,PD84,LS03,AM04,KSS05}. An interesting behavior was found that the ratios of shear viscosity over entropy density ($\eta/s$) for many substances have minimum values at  their corresponding critical temperatures  \cite{RAL07}. By using string theory method, Kovtun-Son-Starinets found that the $\eta/s$ has a limiting value of $\hbar/4\pi$, which is called KSS bound \cite{KSS05}. In relativistic heavy ion collisions, experimental data indicated that the $\eta/s$  of the quark gluon plasma (QGP) is close to the KSS bound, which means that QGP matter is almost perfect fluid \cite{BC05,PC05,SC05,Heinz,Song,Shen,Reining}. So far there are a couple of approaches to calculate shear viscosity \cite{Ma_book}. From theoretical viewpoint,  one can obtain shear viscosity from the Chapman-Enskog and the relaxation time approaches \cite{PD84,LS03,SP12,AW12,XJ13}. From the transport simulation viewpoint, one can extract shear viscosity in the periodic box \cite{AM04,CJ08,JA18}. Also the shear viscosity can be estimated by the mean free path of nucleons ~\cite{DQ14,LiuHL}. What's more, one can extract shear viscosity from the width and energy of the giant dipole resonance (GDR) \cite{NA09,ND11,GuoCQ,DM17,GDR2} or fragment production \cite{SP10} or fitting formula \cite{ZhouCL1,DXG16}.

There are two general methods, namely the Green-Kubo formula and SLLOD algorithm to calculate shear viscosity, which were extensively used for the molecular dynamics simulations \cite{GP05,MM12,ZY15,ZhouCL1,GuoCQ}. One of motivations of the present work is to give a new form of the Green-Kubo formula. By a comparison 
among the traditional Green-Kubo formula, a new form of the Green-Kubo formula is presented and  its validity of those methods is discussed by the  SLLOD algorithm.

The paper is organized as follows: In Sec. \ref{ThreeMethods}, we introduce the simulation model and analysis methods. In Sec. \ref{resultsAA}, we discuss the shear viscosity by different approaches with and without the Pauli blocking at different densities. Conclusion is given in Sec. \ref{summary}.

\section{Nuclear system and analysis methods}
\label{ThreeMethods}

\subsection{ImQMD model}
\label{ImQMDModel}

Generally there are two types of transport models, i.e. Boltzmann-Uehling-Uhlenbeck type and quantum molecular dynamics type for describing  heavy-ion collisions at low and intermediate energies~\cite{GF88,Aichelin,LiBA,Ono,XuJ2,XJ16,ZYX18} and had numerous applications 
\cite{LiBA2,Colonna,Bonasera,MaCW,LiSX,HeWB,Wei,Zhang,Yu,Yan,WangSS,HeYJ}. 
In this work, an improved quantum molecular dynamics (ImQMD) model is utilized \cite{ZYX06}. The potential energy density without spin-orbit term in the ImQMD model reads \cite{ZYX06,WN16}:
\begin{equation}
\begin{split}
V_{loc}&= \frac{\alpha}{2}\frac{\rho^{2}}{\rho_{0}} + \frac{\beta}{\gamma+1}\frac{\rho^{\gamma+1}}{\rho_{0}^{\gamma}} +\frac{g_{sur}}{2\rho_{0}}(\bigtriangledown\rho)^{2}     \\
           &+g_{\tau} \frac{\rho^{\eta +1}}{\rho_{0}^{\eta}} +\frac{g_{sur,iso}}{\rho_{0}}[\nabla(\rho_{n}-\rho_{p})]^2+\frac{C_s}{2\rho_{0}}\rho^{2} \delta^{2},
\end{split}                              
\label{QMDpotential}
\end{equation}
where $\rho, \rho_{n}$ and $\rho_{p}$ are the nucleon, neutron and proton densities, respectively. Here the saturation density of nuclear matter is $\rho_{0} \approx 0.16 fm^{-3}$. And $\delta = (\rho_{n}-\rho_{p})/(\rho_{n}+\rho_{p})$ is the isospin asymmetry. 
For simplicity, we investigate the shear viscosity of infinite nuclear matter without mean field. And a periodic box is constructed within the framework of the ImQMD model as did in Ref.~\cite{ZYX18}. The conditions for box initialization are particle number $A$, density $\rho$ and temperature $T_{0}$. For the simulations, the particle number $A$ is fixed at 600and the total  nucleon-nucleon  cross section ($\sigma_{NN}$) is fixed at 40 mb. Then the box size would be dependent on the nuclear density.  Also we just consider the  symmetric nuclear matter (i.e., $\rho_{n} = \rho_{p}$) here. With the simulations by the ImQMD model, shear viscosity is extracted by different approaches. 

\subsection{Pauli blocking effect on distribution of system}
\label{PauliBlock}
\begin{figure}[htb]
\setlength{\abovecaptionskip}{0pt}
\setlength{\belowcaptionskip}{8pt}
\includegraphics[scale=0.59]{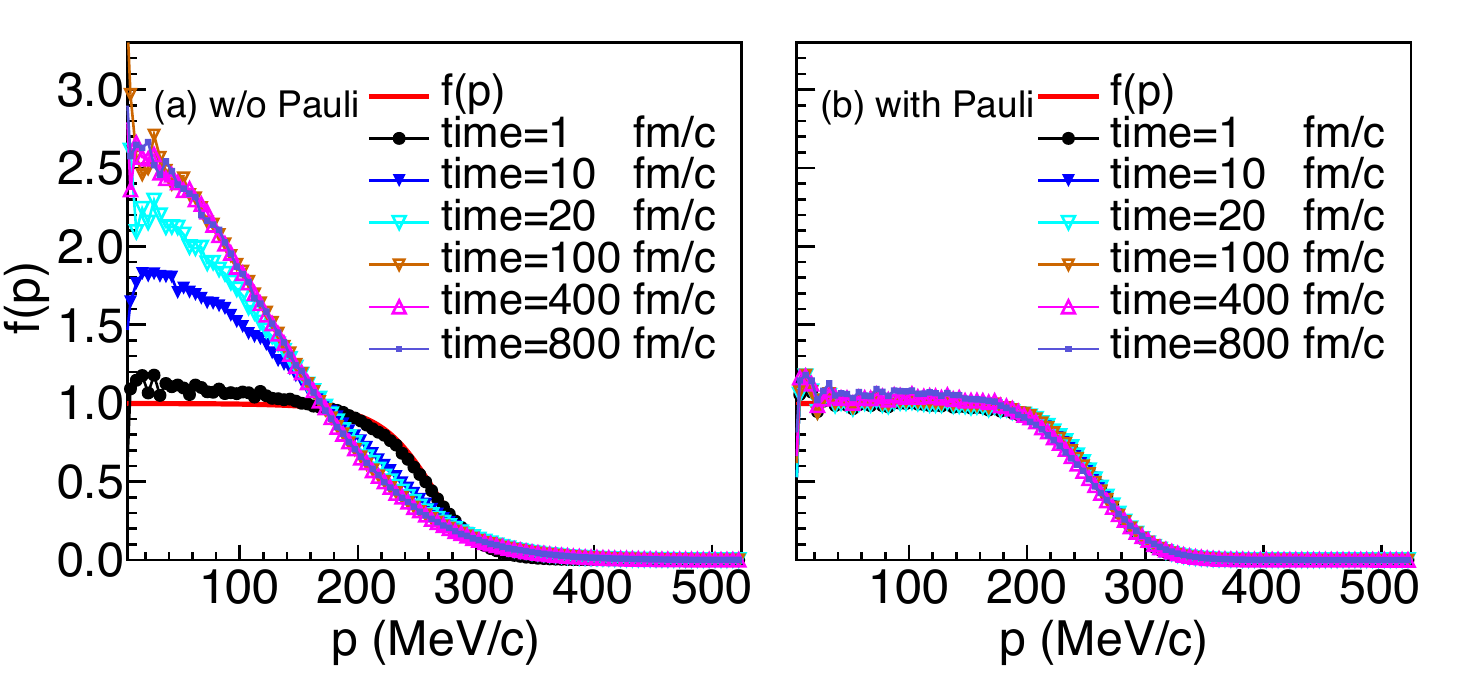}
\caption{(Color online) Time evolution of momentum distributions at density 1.0$\rho_{0}$ and temperature $T$ = 6 MeV. (a) without and (b) with the Pauli blocking.  }
\label{fig:fig1}
\end{figure}

In the present work, we would consider the effects of  Pauli blocking on shear viscosity. However, one thing which is obviously affected by the Pauli blocking is the momentum distribution of the system as shown in Fig.~\ref{fig:fig1}. The Pauli blocking would determine what kind of distribution for a system, eg. either Fermi-Dirac distribution or classical one (Boltzmann distribution).

In this figure, we consider the cases without the Pauli blocking in Fig.~\ref{fig:fig1} (a) as well as with the Pauli blocking in Fig.~\ref{fig:fig1} (b). Here the momenta of particles are initialized by the Fermi-Dirac equation in a condition with  a given set of density ($\rho$), temperature ($T$) and chemical potential ($\mu$). The Fermi-Dirac distribution has the form of 
\begin{equation}
\begin{split}
f(\epsilon)= \frac{1}{\exp(\frac{\epsilon-\mu}{T})+1}, 
\end{split}                              
\label{FermiDiracDis}
\end{equation}
where  $\epsilon$ = $p^{2}/(2m)$ for non-relativistic case and while $\epsilon$ = $\sqrt{p^{2} + m^{2}}$ for relativistic case. The initial momentum distributions are also shown with the red lines in these figures. In Fig.~\ref{fig:fig1}(a) without the Pauli blocking,  the shape of momentum distribution tends to the Boltzmann one as time increases. It is worth to mention that in the standard ImQMD model, the occupation probability in the Pauli blocking algorithm is calculated by a Wigner density in phase space cell \cite{ZYX18}. This method underestimates the occupation probability in the nuclear matter, and results in a larger collision rate than the analytically evaluated one. In this work, we calculate the occupation probability by using the Fermi-Dirac distribution with the calculated density and temperature at each spatial point. As we checked, the collision rate becomes reasonable comparing to the analytical one.

\subsection{Two forms of the Green-Kubo formula}
\label{GreenKubo-for}

\begin{figure}[htb]
\setlength{\abovecaptionskip}{0pt}
\setlength{\belowcaptionskip}{8pt}
\includegraphics[scale=0.56]{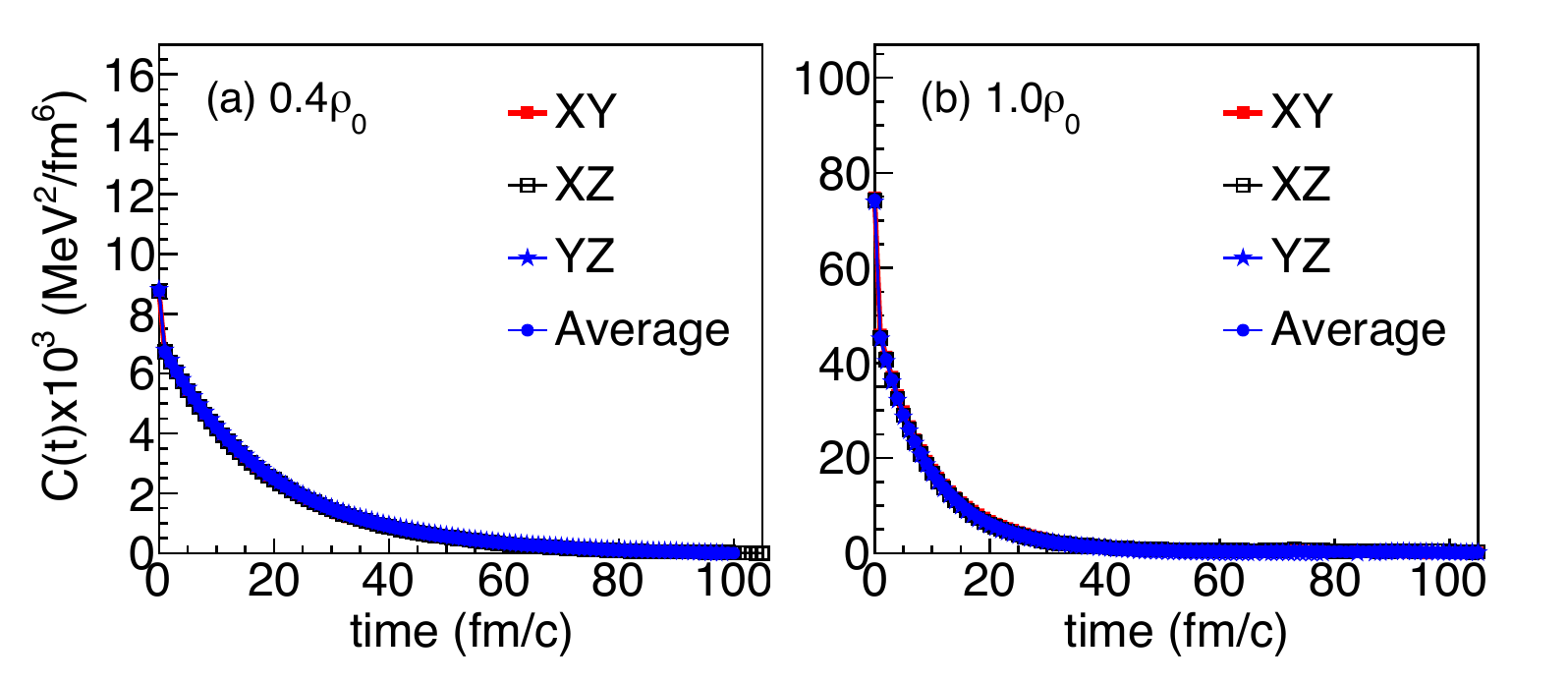}
\caption{(Color online) The evolution of autocorrelation function for different components at  $T_{0}$ = 30 MeV and different densities of $0.4\rho_{0}$ (a) or  $1.0\rho_{0}$ (b)  without mean field.}
\label{fig:fig2}
\end{figure}

One of approaches in this work to calculate shear viscosity is  the Green-Kubo formula which can be derived from the linear response theory. This method was extensively discussed in molecular dynamics simulations \cite{GP05,MM12,ZY15}. The Green-Kubo formula for the calculation of shear viscosity is by an integral of the stress autocorrelation function (SACF) \cite{RK66,DJE08,AH84}:
\begin{equation}
\begin{split}
\eta_{nor} = \frac{V}{T} \int_{t_{0}}^{\infty} C(t) dt,
\end{split}                              
\label{GKubo1}
\end{equation}
where $V$ and $t_{0}$ are system volume and equilibrium time, respectively. The $\eta$ with subscript in Eq.~\ref{GKubo1} means a normal (or traditional) form of the Green-Kubo formula. In last decades, it was extensively used in molecular dynamics simulations. However, it was mostly used for a classical system, not for the Fermi-Dirac distribution system. Thus in this work, we re-derive  and give a new form of the Green-Kubo formula for the calculation of shear viscosity, i.e. 
\begin{equation}
\begin{split}
\eta_{new} = \frac{VNm}{\langle \sum_{i}^{N}p_{ix}^{2} \rangle} \int_{t_{0}}^{\infty} C(t) dt,
\end{split}                              
\label{GKubo1-1}
\end{equation}
where $N$ is total particle number and $m$ is particle mass. The derivation can be found in the Appendix~[\ref{APPA}] and a more general form for shear viscosity as in Eq.~\ref{DRIEQ-52} . One can see that the new form the Green-Kubo formula does not relate to the temperature but instead of the particle momenta. And in both Eq.~\ref{GKubo1} and Eq.~\ref{GKubo1-1}, autocorrelation function $C(t)$ keeps the same and it reads, 
\begin{equation}
\begin{split}
C(t) = \langle P_{\alpha\beta}(t)P_{\alpha\beta}(t_{0}) \rangle {\quad} \alpha, \beta=x, y, z, 
\end{split}                              
\label{GKubo2}
\end{equation}
where $P_{\alpha\beta}(t)$ is off-diagonal element of stress tensor. The bracket $\langle...\rangle$ denotes average over equilibrium ensemble (average by simulation events). Fig.~\ref{fig:fig2} shows the autocorrelation function $C(t)$ as a function of time for different tensor components at densities of $0.4\rho_{0}$ and $1.0\rho_{0}$ at $T_{0}$ = 30 MeV. Here temperature index $T_{0}$ means the initial temperature we set for the box system.
One can see that $C(t)$ tends to zero with increasing time, which indicates that $C(t)$ is convergent. However, $C(t)$ at the lower density has longer correlation time in comparison  with higher one, as shown in Fig.~\ref{fig:fig2}(a) and (b). 

The macroscopic momentum flux in the volume $V$ is given by
\begin{equation}
\begin{split}
P_{\alpha\beta}(t) = \frac{1}{V}\int d^{3}r P_{\alpha\beta}(\vec{r}, t),
\end{split}                              
\label{GKubo3}
\end{equation}
and the local stress tensor is defined as
\begin{equation}
\begin{split}
&P_{\alpha\beta}(\vec{r},t) = \sum_{i}^{N} \frac{p_{{i}\alpha}p_{{i}\beta}}{m_{i}}\rho_{i}(\vec{r},t)   \\
& + \frac{1}{2} \sum_{i}^{N} \sum_{i\neq j}^{N} F_{ij\alpha}R_{ij\beta} \rho_{j}(\vec{r},t)             \\   
 & + \frac{1}{6} \sum_{i}^{N} \sum_{i\neq j}^{N}\sum_{i\neq j \neq k}^{N} (F_{ijk\alpha}R_{ik\beta}+ F_{jik\alpha}R_{jk\beta} )\rho_{k}(\vec{r},t) \\
&+\cdots,
\end{split}                              
\label{GKubo4}
\end{equation}
where $F_{ij}$ and $\vec{R}_{ij} = \vec{r}_{j} - \vec{r}_{i}$ are interaction force and relative position of particle $i$ and $j$. The first term of right-hand side is momentum term, the second one is two-body interaction term, and the third one is three-body interaction term. Actually, the mean field is not considered for our simulations but we put the interaction terms here. Based on the ImQMD model, the $i$-th particle density distribution is given by
\begin{equation}
\begin{split}
\rho_{i}(\vec{r}) = \frac{1}{(2\pi \sigma^{2})^{3/2}} \exp[-\frac{(\vec{r}-\vec{r}_{i})^{2}}{2\sigma^{2}}], 
\end{split}                              
\label{GKubo5}
\end{equation}
where $\sigma$ is the wave-packet width which is taken as 2.0 $fm$ in the present work.

\subsection{The Gaussian thermostated SLLOD algorithm}
\label{GTSLLOD}

Another approach is the SLLOD, which was named by Evans and  Morriss \cite{GP06} and related to the dynamics with (artificial) strain rate $\dot{\gamma}$, algorithm  for non-equilibrium molecular dynamics (NEMD) calculation and has been  extensively applied to predict the rheological properties of real fluids \cite{GP06}. The SLLOD algorithm for shear viscosity was actually applied to a planar Couette flow field at shear rate $\dot{\gamma}$ = $\partial v_{x}/\partial y$ which is the change in streaming velocity $v_{x}$ in the $x$-direction with vertical position $y$. One can get the shear viscosity at shear rate
\begin{equation}
\begin{split}
\eta = -\frac{\langle P_{xy} \rangle }{\dot{\gamma}}.
\end{split}                              
\label{GKubo6}
\end{equation}
With adding shear rate to the system, the dynamical equations of motion of the system are rewritten as
\begin{align}
&\frac{d\vec{r}_{i}}{dt}=\frac{\vec{p}_{i}}{m_{i}} + \dot{\gamma}y_{i} \hat{x}               
\label{GKubo77}  \\
&\frac{d\vec{p}_{i}}{dt}=\vec{F}_{i}-\dot{\gamma}p_{yi} \hat{x}-h \vec{p}_{i}.        
\label{GKubo88}
\end{align}
Eq.~\ref{GKubo77}  and Eq.~\ref{GKubo88} are called the SLLOD equations. In order to keep the kinetic energy conservative, one needs a `thermostat'. Thus a multiplier is applied to motion equations.  Considering conservation of kinetic energy, one can get
\begin{equation}
\begin{split}
h = \frac{\sum_{i}(\vec{F}_{i}\cdot \vec{p}_{i}/m_{i}-\dot{\gamma}p_{xi}p_{yi}/m_{i})}{\sum_{i}p_{i}^{2}/m_{i}}.
\end{split}                              
\label{GKubo9}
\end{equation}
One should notice that the shear rate can not be too large or small.  If it is small, flow field can not be implanted to the system.
\begin{figure}[htb]
\setlength{\abovecaptionskip}{0pt}
\setlength{\belowcaptionskip}{8pt}
\includegraphics[scale=1.1]{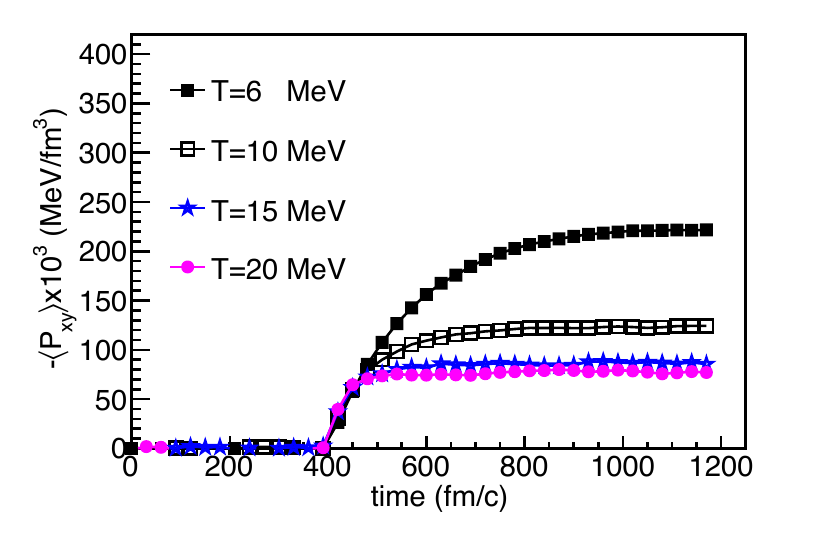}
\caption{(Color online) Stress tensor as a function of time at different temperatures and density of 0.4$\rho_{0}$ without the mean field.}
\label{fig:fig3}
\end{figure}
On the other hand,  when shear rate is too large, energy can not be kept conservative  when  the dynamical equations of motion are solved with a finite time step. As in Fig.~\ref{fig:fig3} shows, shear viscosity decreases with big shear rate $\dot{\gamma}$ due to the energy loss with big values of $\dot{\gamma}$. So in our simulations, different temperatures are correspondent to different values of shear rate. Here $\dot{\gamma}$ = 0.0003 $c/fm$, $\dot{\gamma}$ = 0.0005 $c/fm$ and $\dot{\gamma}$ = 0.002 $c/fm$ are taken for  $T_{0}$ = 4 MeV, $T_{0}$ = 6 MeV and $T_{0} \geq  $10 MeV, respectively.

\begin{figure*}
\setlength{\abovecaptionskip}{0pt}
\setlength{\belowcaptionskip}{8pt}
\includegraphics[scale=1.1]{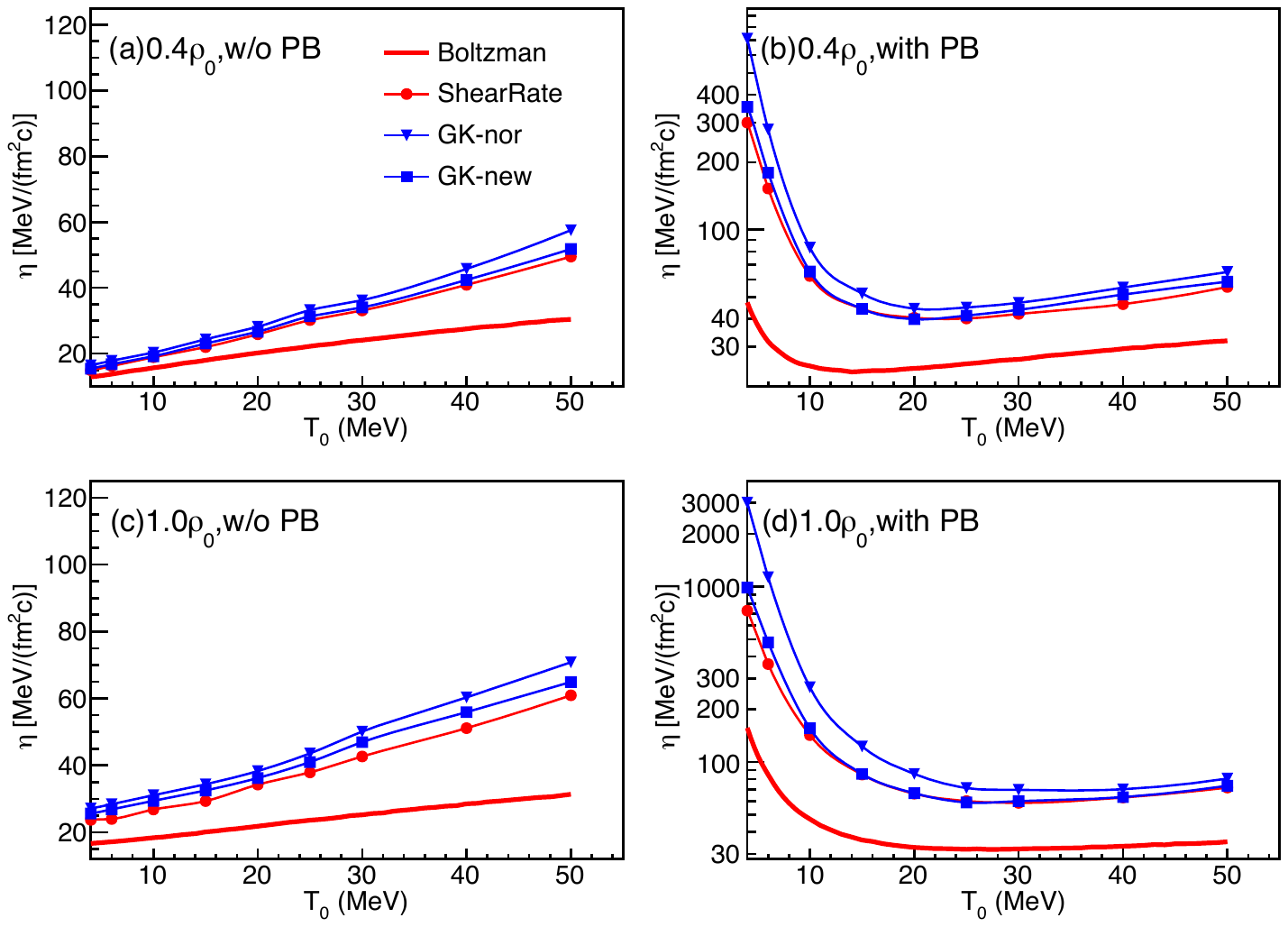}
\caption{(Color online) Different calculations of shear viscosity as a function of temperature in cases of w/ or w/o Pauli blocking (PB) at  0.4 $\rho_0$ and 1.0$\rho_0$.  }
\label{fig:fig4}
\end{figure*}

\begin{figure}[htb]
\setlength{\abovecaptionskip}{0pt}
\setlength{\belowcaptionskip}{8pt}
\includegraphics[scale=1.1]{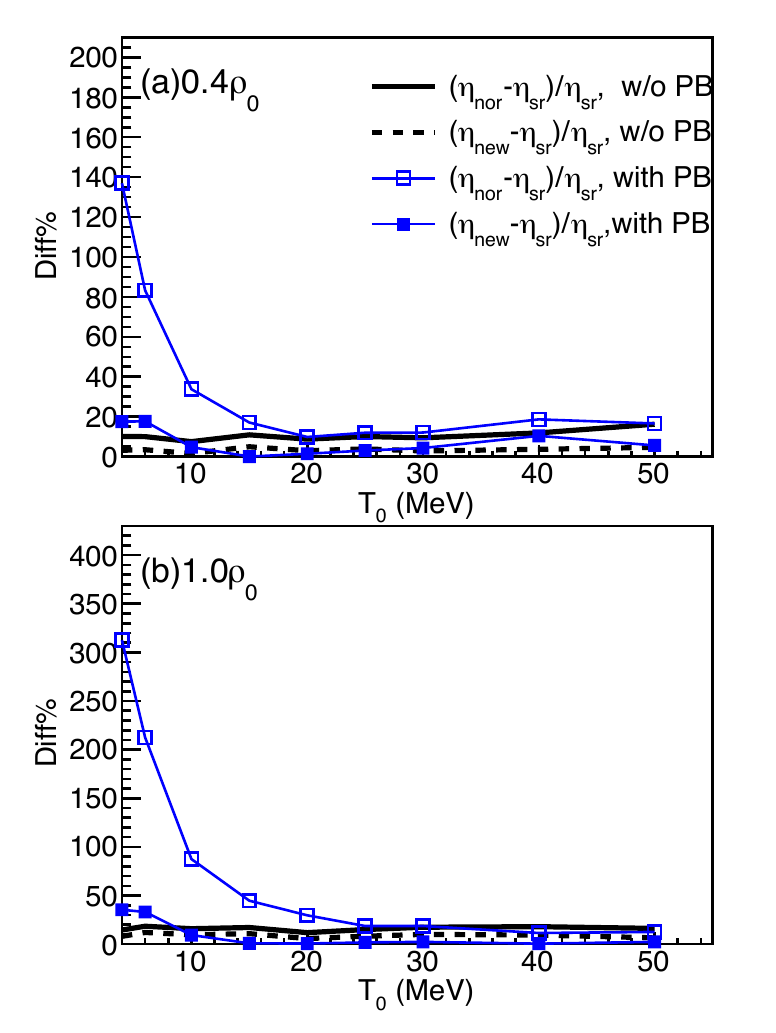}
\caption{(Color online) The respective relative difference between the shear rate method and the normal Green-Kubo formula or and the new Green-Kubo formula as a function of temperature.}
\label{fig:fig5}
\end{figure}

\subsection{The linear Boltzmann equation set}
\label{BoltzmannEquation}
How good are they for these methods mentioned above while they are applied  to the nuclear matter? As a comparison, a linear Boltzmann equation set which can be used to calculate the shear viscosity of uniform nuclear matter is presented here. One can get the expression by solving Boltzmann equation \cite{PD84,LS03,BB19}:
\begin{equation}
\begin{split}
\eta& = \frac{5T}{9}\frac{\Big{(}\int d^{3}p{\,\,}p^{2}f(p)\Big{)}^{2}}{\int d^{3}p_{1}d^{3}p_{2}d\Omega {\,} v_{12}q_{12}^{4}sin^{2}{\theta}\frac{d\sigma_{NN}}{d\Omega} f_{1}f_{2}\tilde{f}_{3}\tilde{f}_{4}}, 
\end{split}                              
\label{BoltzmannSet}
\end{equation}
where $v_{12} = |\vec{v}_{1}-\vec{v}_{2}|$ and $q_{12} = |\vec{p}_{1}-\vec{p}_{2}|/2$ are relative velocity and relative momentum, respectively,  $\sigma_{NN}$ is total nucleon-nucleon cross section, and $\tilde{f}_{3}$$\tilde{f}_{4}$ = [1-$f(p_{3})$][1-$f(p_{4})$] is the Pauli blocking term. In Eq.~\ref{BoltzmannSet}, the conservation of momentum and energy between $p_{1}, p_{2}$ and $p_{3}, p_{4}$ are needed to be taken into account. 

\section{Shear viscosity with different approaches}
\label{resultsAA}

By using different approaches as we mentioned above, as shown in Fig.\ref{fig:fig4},  shear viscosity is calculated at different temperatures and densities with and without the Pauli blocking. One sees that in Fig.~\ref{fig:fig4}(a) and Fig.~\ref{fig:fig4}(c) without the Pauli blocking, shear viscosity increases with increasing temperature at both densities of $0.4\rho_{0}$ and $1.0\rho_{0}$.  As temperature increases, exchanging of momentum among the particles, a `stopping' effect increases between two flow layers. That indicates that shear viscosity would increase. Seeing from  Fig.~\ref{fig:fig4} (b) and Fig.~\ref{fig:fig4}(d) with the Pauli blocking, unlike Fig.~\ref{fig:fig4}(a) and Fig.~\ref{fig:fig4}(c) in the low temperature region, shear viscosity increases with decreasing temperature. It is due to a stronger Pauli blocking effect  in the low temperature region. At lower temperature and with  stronger Pauli blocking effect, it indicates that a good Fermi sphere is formed, also energy and momentum transport  becomes quite efficient \cite{YK16}. As the temperature increases, the Pauli blocking effect reduces, and shear viscosity decreases. And as temperature increases again, the collision number would increase, so shear viscosity increases. Comparing among these three approaches, we find that shear viscosity determined by the Boltzmann type equation is lower than those which are calculated by the  shear rate and the Green-Kubo method. At very low densities such as 0.05$\rho_0$, however, we found that the result by the Boltzmann type equation (Eq.~\ref{BoltzmannSet}) is consistent with other methods, which indicates that low density approximation for Eq.~\ref{BoltzmannSet} could be valid, then one would expect that shear viscosities determined by the shear rate and the Green-Kubo approaches are the same. As displayed in  Fig.~\ref{fig:fig4} (a) and Fig.~\ref{fig:fig4} (c) without the Pauli blocking, shear viscosities from the shear rate method (the SLLOD algorithm), the normal form of the Green-Kubo formula, and the new form of the Green-Kubo formula are consistent with each other.  However, as the Pauli blocking is taken into account as shown in  Fig.~\ref{fig:fig4} (b) and Fig.~\ref{fig:fig4} (d), one sees that in low temperature region, the shear viscosity from the normal form of the Green-Kubo formula is  higher than those from both the SLLOD algorithm and the new form of the Green-Kubo formula. However, shear viscosities from the SLLOD algorithm and the new form of the Green-Kubo formula are almost the same. The relative difference  is shown in Fig.~\ref{fig:fig5}. One can find that there is more different between the normal form of the Green-Kubo formula and the SLLOD algorithm or the new form of the Green-Kubo formula at lower temperatures. It is obviously that the standard GK formula leads to larger shear viscosity in fermionic system especially at low temperatures. 

\section{Conclusions}
\label{summary}
In summary, shear viscosities are obtained by the SLLOD algorithm and the Green-Kubo formula in the framework of the ImQMD simulation and are compared with the Boltzmann equation method in the present work. By comparisons among different calculation methods, it is found that shear viscosity with the Boltzmann equation method is less than those from the SLLOD algorithm and the Green-Kubo formula. More interestingly,  a new form of the Green-Kubo formula (Eq.~\ref{GKubo1-1}) is presented for shear viscosity calculation. By comparison with the SLLOD algorithm, we found that the standard GK formula leads to larger shear viscosity in fermionic systems especially at low temperatures. And the new form of  the Green-Kubo formula for shear viscosity is consistent with the SLLOD algorithm for both classical and non-classical systems.

\begin{acknowledgments}

Thanks for helpful discussions with P. Danielewicz and H. Lin. This work was partially supported by the National Natural Science Foundation of China under Contract Nos. 11947217, 11890710 and 11890714, China Postdoctoral Science Foundation Grant  No. 2019M661332, Postdoctoral Innovative Talent Program of China No. BX20200098, the Strategic Priority Research Program of the CAS under Grants No. XDB34000000,  the 
Guangdong Major Project of Basic and Applied Basic Research No. 2020B0301030008.

\end{acknowledgments}

\begin{appendix}
\section{The Green-Kubo formula for shear viscosity}
\label{APPA}

The particle density can be written as (for simplicity, here we use $\delta$-function to replace the Gaussian wave-packet). And the derivation of the Green-Kubo formula for shear viscosity is tedious. Here we give a simple introduction which is based on Ref.~\cite{DJE08}. For  more details one can find in Refs.~\cite{DJE96,DJE08}.

\subsection{$\vec{\bf{r}}$ and $\vec{\bf{k}}$-space representations}

The streaming velocity, mass density, momentum density and stress tensor at position $\vec{\bf{r}}$ and time $t$ can be written:
\begin{align}
\label{DRIEQ-1}
&{\bf u} (\vec{\bf{r}},t)=\frac {\sum_{i}^{N} m_{i} {\dot{\vec{\bf{r}}}_{\it{i}}} \delta (\vec{\bf{r}}-\vec{\bf{r}}_{i})}{\sum_{i}^{N} m_{i} \delta (\vec{\bf{r}}-\vec{\bf{r}}_{i})},           \\               
\label{DRIEQ-2}
&\rho(\vec{\bf{r}},t)=\sum_{i}^{N} m_{i} \delta (\vec{\bf{r}}-\vec{\bf{r}}_{i}),\\
\label{DRIEQ-3}
&J(\vec{\bf{r}},t)=\rho(\vec{\bf{r}},t){\bf u} (\vec{\bf{r}},t)=\sum_{i}^{N} m_{i} {\dot{\vec{\bf{r}}}_{\it{i}}} \delta (\vec{\bf{r}}-\vec{\bf{r}}_{i}), \\
\label{DRIEQ-4}
&P_{\alpha\beta}(\vec{\bf{r}},t)=\sum_{i}^{N} \frac{p_{{i}\alpha}p_{{i}\beta}}{m_{i}} \delta (\vec{\bf{r}}-\vec{\bf{r}}_{i})    \notag \\
&+\frac{1}{2} \sum_{i}^{N} \sum_{i\neq j}^{N} F_{ij\alpha}R_{ij\beta} \delta (\vec{\bf{r}}-\vec{\bf{r}}_{j}) \\ \notag
&+\cdots ,
\end{align}
where `$i$' and `$j$'  are indexes of particles and $N$ is particle number. For Eq.~(\ref{DRIEQ-4}), the momentum density conservation law
\begin{align}
\label{DRIEQ-1-1}
\frac{\partial J(\vec{\bf{r}},t)}{\partial t } = -\nabla_{\bf r} \cdot \vec{P},
\end{align}
is needed. Moreover it needs
\begin{align}
\label{DRIEQ-1-2}
&\frac{\partial}{\partial \vec{\bf{r}}_{\it{i}} } \delta (\vec{\bf{r}}-\vec{\bf{r}}_{i})= - \frac{\partial}{\partial \vec{\bf{r}} } \delta (\vec{\bf{r}}-\vec{\bf{r}}_{i}),  \\               
\label{DRIEQ-1-3}
&\delta (\vec{\bf{r}}-\vec{\bf{r}}_{i}) -\delta (\vec{\bf{r}}-\vec{\bf{r}}_{j})\approx \vec{R}_{ij}\frac{\partial}{\partial \vec{\bf{r}} } \delta (\vec{\bf{r}}-\vec{\bf{r}}_{j}).
\end{align}

One can define the Fourier transform and inverse in three-dimensions by
\begin{align}
\label{DRIEQ-5}
&f(\vec{\bf{k}}) = \int d^{3}{\bf r} f(\vec{\bf{r}})  {\rm exp} [ {\rm i} \vec{{\bf k}}\cdot \vec{\bf r}], \\
\label{DRIEQ-6}
&f(\vec{\bf{r}}) = \frac{1}{(2\pi)^{3}}\int d^{3}{\bf k} f(\vec{\bf{k}})  {\rm exp} [- {\rm i}\vec{\bf{k}}\cdot \vec{\bf{r}} ],
\end{align}
where `${\rm i}$' is the unit of imaginary. Then the mass density, momentum density and stress tensor in $\vec{\bf{k}}$-space are given
\begin{align}
\label{DRIEQ-7}
&\rho(\vec{\bf{k}},t)={\sum_{i}^{N}} {m_{i}} {\rm exp} [ {\rm i} \vec{{\bf k}}\cdot \vec{\bf r}_{i} ],  \\
\label{DRIEQ-8}
&J(\vec{\bf{k}},t)=\sum_{i}^{N} m_{i} {\dot{\vec{\bf{r}}}_{\it{i}}} {\rm exp} [ {\rm i} \vec{{\bf k}}\cdot \vec{\bf r}_{i} ], \\
\label{DRIEQ-9}
&P_{\alpha\beta}(\vec{\bf{k}},t)=\sum_{i}^{N} \frac{p_{{i}\alpha}p_{{i}\beta}}{m_{i}} {\rm exp} [ {\rm i} \vec{{\bf k}}\cdot \vec{\bf r}_{i} ] \\ \notag
&+\frac{1}{2} \sum_{i}^{N} \sum_{i\neq j}^{N} F_{ij\alpha}R_{ij\beta} {\rm exp} [ {\rm i} \vec{{\bf k}}\cdot \vec{\bf r}_{j} ]+\cdots \; .
\end{align}

\subsection{Shear viscosity and strain rate}

The stress tensor corresponds to the strain rate (for simplicity, only the $x-y$ component is considered) reads
\begin{equation}
\begin{split}
P_{xy}=-\eta \gamma(t),
\end{split}                              
\label{DRIEQ-10}
\end{equation}
where $\eta$ is static shear viscosity. Also one can see it in Eq.~(\ref{GKubo6}) but the strain rate here is time dependent. The most general linear relation between the strain rate and the shear stress can be written in the time domain as
\begin{equation}
\begin{split}
P_{xy}(t)=-\int_{0}^{t} ds \; \eta_{M}(t-s)\gamma(s),
\end{split}                              
\label{DRIEQ-11} 
\end{equation}
where $\eta_{M}(t)$ is called the Maxwell memory function. The memory function explains that the shear stress at time $t$ is not simply linearly proportional to the strain rate at the current time $t$, but to the entire strain rate process, over times $0 \leqslant s \leqslant t$. For the frequency dependent Maxwell viscosity is
\begin{equation}
\begin{split}
\tilde{\eta}_{M}(\omega)= \frac{\eta}{1+{\rm i} \omega \tau_{M}},
\end{split}                              
\label{DRIEQ-12} 
\end{equation} 
where $ \tau_{M}$ is the Maxwell relaxation time which controls the transition frequency between low frequency viscous behaviour and  high frequency elastic behavior. In Eq.\ref{DRIEQ-12}, $\tilde{\eta}_{M}$ is the Fourier-Laplace transform which is read as
\begin{equation}
\begin{split}
\tilde{\eta}(\omega)= \int_{0}^{\infty} dt\; {\rm exp}[-{\rm i}\omega t] \eta(t).
\end{split}                              
\label{DRIEQ-13} 
\end{equation}

\subsection{Shear viscosity of the Green-Kubo formula}

We separate vector-dependent momentum density into longitudinal (${\bf J}^{||}$) and transverse (${\bf J}^{\bot}$) parts.  Considering a transverse momentum density ${\bf J}^{\bot}$($\vec{\bf k}$,t), for simplicity, we define the coordinate system in which  $\vec{\bf k}$ is in $y$-direction and ${\bf J}^{\bot}$ is in the $x$-direction:
\begin{align}
\label{DRIEQ-28} 
J_{x}(k_{y},t) = \sum_{i}mv_{xi}(t){\rm exp}[{\rm i}k_{y}y_{i}(t)].
\end{align}
According to Eq.~(\ref{DRIEQ-9}), one can get
\begin{align}
\label{DRIEQ-29} 
\dot{J}_{x}(k_{y},t) = {\rm i}k_{y}P_{xy}(k_{y},t).
\end{align}
In Ref.~\cite{DJE08}, by Mori-Zwanzig formalism, one can get the shear viscosity $\eta (t)$ which is time-dependent, i.e. 
\begin{align}
\label{DRIEQ-51} 
\eta(t) = \frac{VNm}{\langle J_{x}(k_{y})J_{x}^{\ast}(k_{y})\rangle}\langle P_{xy}(t)P_{xy}(0) \rangle.
\end{align}
By the Fourier-Laplace transform of $\eta(t)$, one gets
\begin{align}
\label{DRIEQ-51} 
\tilde{\eta}(\omega) = \frac{VNm}{\langle J_{x}(k_{y}=0)J_{x}^{\ast}(k_{y}=0)\rangle} \\
\times \int_{0}^{\infty}\langle P_{xy}(t)P_{xy}(0) \rangle  {\rm exp}[-{\rm i}\omega t]dt.   \notag
\end{align}
As in Eq.~(\ref{DRIEQ-12}), static shear viscosity needs $\omega \rightarrow$0. Then one gets
\begin{align}
\label{DRIEQ-52} 
\eta = \frac{VNm}{\langle J_{x}(k_{y} = 0) J_{x}^{\ast}(k_{y} = 0)\rangle} \int_{0}^{\infty}\langle P_{xy}(t)P_{xy}(0) \rangle dt.
\end{align}
Here it should be noticed
\begin{align}
\label{DRIEQ-53-0} 
P_{xy}(t) = \lim_{k_{y}\rightarrow0}\frac{P_{xy}(k_{y},t) }{V}, \\
\label{DRIEQ-53-1} 
P_{xy}(0) = \lim_{k_{y}\rightarrow0}\frac{P_{xy}(k_{y},0) }{V}, 
\end{align}
where $V$ is system volume. For the norm of the transverse current, one can get
\begin{align}
\label{DRIEQ-54} 
&\langle J_{x}(k_{y}=0) J_{x}^{\ast}(k_{y}=0)\rangle \\
&=\langle \sum_{i}^{N}p_{xi} \sum_{j}^{N}p_{xj}\rangle \notag\\
&=\langle \sum_{i}^{N}p_{xi}^{2}\rangle +N(N-1)\langle p_{1x}p_{2x}\rangle.  \notag
\end{align}
At equilibrium, $ p_{1x}$ is independent of  $p_{2x}$, so the second term of right-hand side of Eq.~(\ref{DRIEQ-54}) is zero. Then we can obtain a new form of the Green-Kubo formula for shear viscosity
\begin{align}
\label{DRIEQ-55} 
\eta_{new} = \frac{VNm}{\langle \sum_{i}^{N}p_{xi}^{2}\rangle} \int_{0}^{\infty}\langle P_{xy}(t)P_{xy}(0) \rangle dt,   
\end{align}
where $\langle \cdots \rangle$ denotes ensemble average. For an equilibrium system which obeys the Boltzmann distribution, one can get
\begin{align}
\label{DRIEQ-56} 
\langle \sum_{i}^{N}p_{xi}^{2} \rangle = \langle \frac{1}{3}\sum_{i}^{N}p_{i}^{2} \rangle = Nmk_{B}T, 
\end{align}
where $T$ is temperature and $k_{B}$ ($k_{B}$ = 1) is the Boltzmann constant. Then the normal Green-Kubo formula for shear viscosity can be given
\begin{align}
\label{DRIEQ-55} 
\eta_{nor} = \frac{V}{T} \int_{0}^{\infty}\langle P_{xy}(t)P_{xy}(0) \rangle dt.    
\end{align}

\end{appendix}

\end{CJK*}
\end{document}